# Velocity Map Imaging and Cross Sections of Fe(CO)$_5$ for FEBIP Applications

Maria Pintea[1], Nigel Mason[1,2], Maria Tudorovskaya[3]

[1]Kent University, Department of Molecular Processing, School of Physical Sciences, CT2 7HZ, Canterbury, UK

[2]School of Physical Sciences, The Open University, MK7 6AA, Milton Keynes, UK

[3]Quantemol Ltd, London, UK

**Abstract**

The present paper intends to be a new study of a widely used precursor in nanostructure deposition and FEBID processes with focus on its fragmentation at collisions with low energy electrons. Newer developments in nanotechnology with applications to Focused Electron Beam Induced Deposition (FEBID) and Extreme Ultraviolet Lithography (EUVL), based on irradiation induced chemistry come with advances in the size of the nanostructures at the surface and their flexibility in creating highly complex 3D structures. The deformation in the main structures of the FEBID process characterized by elongation, reduction in diameter of the main structure and the deposition of additional thin layers around the structure on the substrate are results of the secondary electrons effect, colliding with energies lower than 20eV. Fe(CO)$_5$ is one of the most used compounds in FEBID processes as it has a high pressure and has been shown to provide high purity deposits (over 90%). This paper combines experiment and simulations to study electron scattering from Fe(CO)$_5$, while experimental data on dissociative electron attachment is presented using the Velocity Slice Map Imaging (VMI) technique that combined with data collected on the CLUSTER apparatus at Comenius University, Bratislava and Quantemol-N simulations presents the fragmentation pathways and channel distribution for each of the resulting negative ions at low energies.

The Quantemol-N simulation package is used to study collision processes of low-energy electrons with Fe(CO)$_5$ molecules including elastic, electronic excitation, and dissociative electron attachment (DEA) cross sections.

## 1. Introduction
### 1.1 Focused electron beam induced dissociation

Focused Electron Beam Induced Deposition (FEBID) [1] is an emerging technique that can be used for nanosize structure growth [2]. As one of the newest methods to write sub-nm level structures, nanorods, nanowires (for tips used in deposition, in mask repair) and in the development of nanomaterials such as superconductors, magnetic materials, multilayer structures, metamaterials [3], [4], it is still in the initial phase. The feed stock gases used in FEBID are metallic compounds containing organic parts such as oxides, carbonyls and halogens. Such gases are dissociated by high energy electrons to produce metallic deposits with current deposits purity rates in Fe of > 95% Lukasczyk et al [5], Fe > 80% in the work of Gavanin et al [3] and Pt with purity > 99% Thorman, Kumar and Ingólfsson [6]. However, in the FEBID process, backscattered, or secondary electrons are produced at low energies that subsequently interact with the target molecule at energies as low as 0 eV. At such low energies, the dominant process leading to the molecule dissociation is Dissociative Electron

Attachment (DEA).

As part of a larger project, ELENA, and a continuation of a chain of projects involved with the study of Focused Electron Beam Induced Deposition (FEBID) and Dissociative Electron Attachment (DEA), CELINA [7], [8] and ARGENT [9], [10], the present study provides a detailed analysis of $Fe(CO)_5$ using Velocity Slice Map Imaging (VMI) whilst also using the Quantemol-N code to generate a set of low energy cross sections. Up to now, no VMI data has been declared on $Fe(CO)_5$ compound, though mass spectrum measurements have been carried out [11], [12], [13] and data can be found on DEA to $Fe(CO)_5$ clusters, degradation of $Fe(CO)_5$ in electron stimulated studies, $Fe(CO)_5$ deposited on Argon nanoparticles, ligand exchange dynamics of $Fe(CO)_5$ in solutions, $Fe(CO)_5$ bond acceptor - donator reactions with X = F, Cl, Br, dynamics studies of $Fe(CO)_5$ by transient ionization and dissociation after ligand stabilization of $Fe(CO)_5$ aggregates [14], [15].

The Velocity Slice Map Imaging (VMI) method uses a time-of-flight spectrometer for mass selection of the ions produced from the interaction of a molecular beam with an electron beam created by an electron gun equipped with a tungsten filament. The product ions are directed towards a MCP detector and a phosphor screen where a 2D picture image is captured by the use of a CCD camera. Used to study the dynamics of dissociative ionization and dissociative electron attachment of CO, $H_2O$, $NH_3$, $CH_4$, $H_2S$, $H_2D_2$, $N_2O$, $CF_4$, $C_2H_2$, the VMI method gives accurate information on angular distribution and kinetic energies of these two processes, however, to date, the method has not been largely used for metal-containing compounds in nanotechnology applications.

This paper is organized as follows: firstly, we discuss the properties of the molecule (molecular geometry, vibrational frequencies and electron affinities) according to the fragmentation channel, then we discuss the experimental setup and the experimental results, while in the last section, we present the computational details and computational results.

**1.2 Molecular target**

Capable of providing iron nanostructures with high purity and exhibiting high ferromagnetic behavior, $Fe(CO)_5$, is a good candidate for the creation of structures at the nanoscale for magnetic sensing elements, hard disc data storage, logic devices, and sensors. Extensive studies on the applicability of Fe-carbonyls to nanosize structure growth by focused electron beam induced dissociation have been carried out, summarized in the work of De Teresa and Pacheco [1] and van Dorp and Hagen [2]. Analysis of the molecular structure of $Fe(CO)_5$ and simulations run using Quantemol-N offer reliable output cross sections values to induced collisions with excited electrons. The geometry of the molecule is presented in Figure 1. The molecule is a symmetric dihedral ($D_{3h}$ point work group) with the Fe atom in the center. The cartesian coordinates used in this work are listed in Table 1. Table 2 summarizes the vibrational frequencies corresponding to the specific fragmentation channels, as well as the electron affinities.

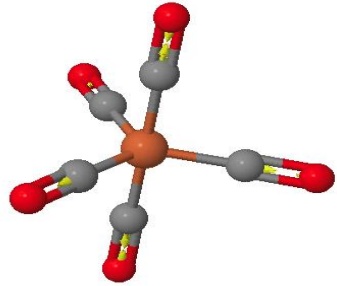

Figure 1. Fe(CO)$_5$ geometry. Orange: Fe atom. Grey: C atoms. Red: O atoms.

**1.3 Electron impact processes**

In this paper we seek to provide the electron impact cross sections for a variety of collision processes including elastic scattering:

$$Fe(CO)_5(X) + e \rightarrow Fe(CO)_5(X) + e, \qquad (1)$$

where X is the ground state denotes the molecule.

Electron-impact excitation:

$$Fe(CO)_5(^1A1) + e \rightarrow Fe(CO)_5(Y) + e, \qquad (2)$$

where Y stands for an excited state different from the ground state and

Dissociative electron attachment:

$$Fe(CO)_5(^1A1) + e \rightarrow Fe(CO)_5^- \rightarrow (CO)_n + Fe(CO)_{5-n}^- \qquad (3)$$

The latter process goes through formation of a resonance, i.e. an incoming electron forming a short-living quasi-bound state with the target molecule. This intermediate resonance then decays into fragments, therefore, the index n=(1...5).

The probability of a scattering process is described with the use of scattering cross-section. Such cross-sections may be computed with the help of the R-matrix theory. The idea of the method lies in separating the scattering space into an inner and outer region. The inner region, separated from the outer region by a sphere of a certain radius, must contain the molecule with its N electrons and the incoming electron which becomes indistinguishable from the molecular orbitals. The outer region only contains the incoming electron which interacts with the potential of the N-electron molecule. The Schrödinger equation is solved separately for the target molecule, in the inner region, and in the outer region, and the solutions are matched at the boundary. For (1) and (2), the cross-section can be derived unambiguously if the wave function of the system is known. In the close-coupling (CC) expansion used in this paper, the wave function in the inner region is presented in terms of N-electron target states, $\Phi_i^N(\mathbf{x_1}...\mathbf{x_N})$, continuum orbitals $u_{ij}(\mathbf{x_{N+1}})$, and quadratically integral functions $\Psi_i^{N+1}$ constructed from target occupied and virtual orbitals:

$$\Psi_k^{N+1} = A \sum_{ij} a_{ijk} \Phi_i^N(\mathbf{x_1}...\mathbf{x_N}) u_{ij}(\mathbf{x_{N+1}}) + \sum_i b_{ij} \Psi_i^{N+1} \qquad (4)$$

In the calculation, some electrons are "frozen" whilst the remaining electrons as well as the scattered electron can be redistributed between remaining bound orbitals, and the scattered electron occupies the continuum orbitals.

For (3), additional data is needed. The R-matrix routines are used to find the energy, E, and the width, Γ, of the quasi-bound states, or resonances, when the incoming electron is temporarily bound to the molecule. It is assumed that the cross-section of scattering via an individual resonance behaves as described by Breit and Wigner:

$$\sigma_{BW}(E, r) = 2\pi \frac{\Gamma^2/4}{(E - V_r(r))^2 - \Gamma^2/4}, \qquad (5)$$

where r is the distance between the dissociating fragments and $V_r$ is the effective resonance potential. It is then assumed that this intermediate state dissociates into fragments effectively behaving like a diatomic molecule. A dissociative electron attachment estimator is incorporated in the Quantemol-N. The model is based on a few assumptions:

- Relative to the active bond length(s), the molecular target potential can be approximated as Morse potential; the negative ion potential is either Morse potential or an exponentially decaying function. This allows vibrational wave functions corresponding to perturbing the bond between the dissociating fragments to be determined.
- The potentials can be described with the vibrational frequency corresponding to the ground vibrational state and the dissociation energy of the molecular target.
- The cross-section has a Breit-Wigner shape; each resonance formed during the scattering process is characterized by its position and width.

More detailed description can be found in the paper of Munro et al [16].

## 2. Experiment
### 2.1 Experimental Setup

The VMI apparatus is presented in Figure 2. The vacuum system is maintained at a pressure of ~$10^{-9}$ mbar by a system of pumps, consisting of a turbopump together with a backing pump, an oil-free Varian TriScroll high-speed scroll pump and a set of ion and Pirani gauges. The gas line is served by Oerlikon Leybold 30 l/s turbo pump at a pressure of $10^{-3}$ mbar. The electrons involved in the electron-molecule collision processes are produced by thermionic emission from the 0.2mm tungsten filament in a the electron gun and collimated with both electrostatic lenses and a magnetic field generated by two Helmholtz coils. The molecular beam is introduced to the chamber through the gas line and it is perpendicular to the electron beam.

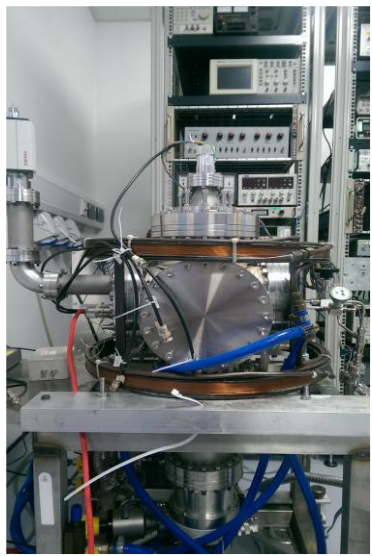
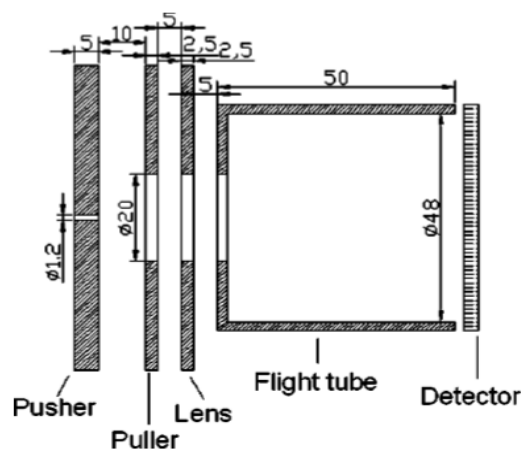

Figure 2. Velocity Slice Map Imaging apparatus at University of Kent, consisting of the high vacuum chamber, electron gun, and Helmholtz coils

## 2.2 Experimental Results and Discussion

In the DEA process $Fe(CO)_5$ undergoes dissociation into five negative ions. The process is characterized by full dissociation as a result of the loss of five ligands to CO. The negative ion mass spectrum with the yields for each of the resonances, taken at Comenius University is presented in Figure 3 and are summarized in Table 2.
The spectrum is in good agreement with the data reported from our TOF VMI (Time – of – Flight Velocity Slice Map Imaging) and with the initial work of Allan et al [17] and Lengyel et al [18].
The possible fragmentation pathways in the collision or ionization process of $Fe(CO)_5$ are:

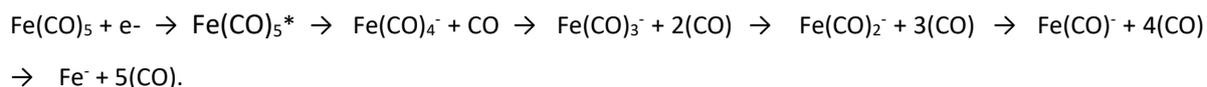

$Fe(CO)_5 + e^- \rightarrow Fe(CO)_5^* \rightarrow Fe(CO)_4^- + CO \rightarrow Fe(CO)_3^- + 2(CO) \rightarrow Fe(CO)_2^- + 3(CO) \rightarrow Fe(CO)^- + 4(CO) \rightarrow Fe^- + 5(CO)$.

The most dominant ion produced by DEA is $Fe(CO)_4^-$ peaking around 0.2eV. Allan et al [17] in their experimental work positioned their $Fe(CO)_4^-$ resonance at a value of 0.08 eV, defined by a narrow peak width of 70eV and a second resonance of the same ion but presenting lower amplitude at 0.26 eV, characterized by a beam resolution of 100meV. The high count rates for the dissociation resonances peaking near 0 eV show that the dominant process at these energies is intramolecular vibration causing the breakage of the ligand and a high rise in the dissociation cross-sections. The loss of one ligand in the DEA process at energies close to 0 eV is considered a transition process of the Fe(CO)$_5$ in the neutral ground state σ to π* excited state, losing one of the axial CO bonds to metal, forming Fe(CO)$_4$-.

The electron affinity of the ions formed in the photo-dissociation process is presented by Engelking and Lineberger [19]. The authors experimentally determined the values by using an electrical discharge ion source and a Wien filter with an Ar ion laser. The values are presented in Table 1, comparable with the work of Shuman et al [11]. The Fe atom affinity is experimentally determined to be 0.164 ± 0.02eV, higher compared to the work of Chen et al [20] exhibiting a value of 0.153eV with the slow electron velocity map imaging (SEVI) method. The thermochemistry data from Shuman et al [11] falls back to the measurements of Engelking and Lineberger [19] with changes to the uncertainty corrections. They employed variable electron and neutral density attachment spectroscopy (VENDAMS) technique and flowing afterglow Langmuir probe mass spectrometer (FALP), to determine the accurate values of electron affinity and bond dissociation energies for the negative ions formed in the DEA process of Fe(CO)$_5$.

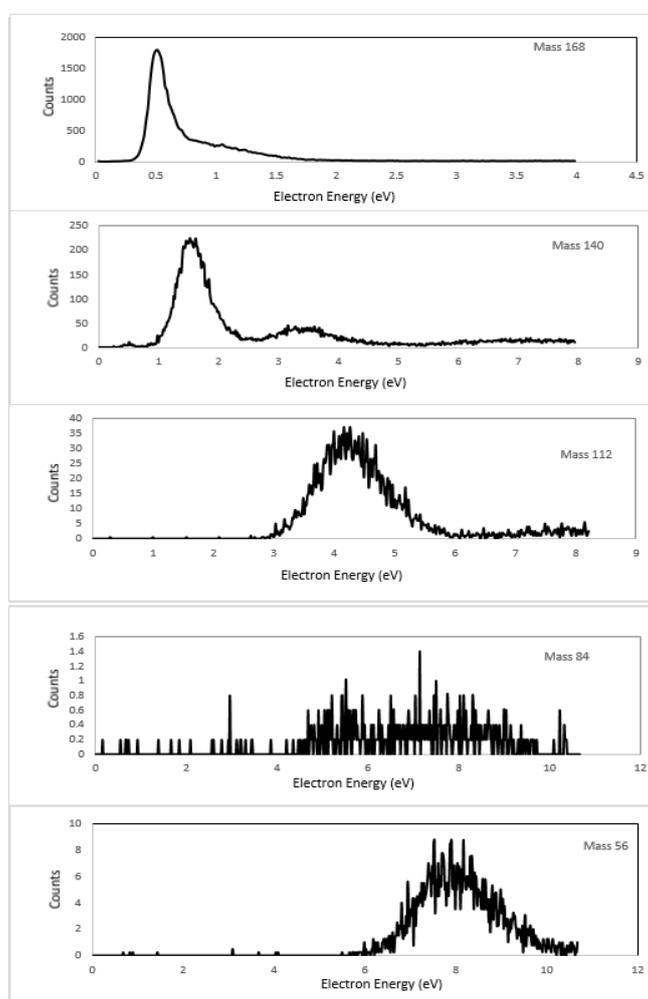

Figure 3. Mass spectrum of Fe(CO)$_5$ recorded at Comenius University, Bratislava

The present kinetic energies have been calculated using the excess energy Ee = EA - BDE + Ei [21], where EA is the electron affinity, BDE is the bond dissociation energy and Ei is the incident electron energy. The maximum kinetic energy for the $Fe(CO)_4^-$ fragment is determined using the electron affinity of the fragment from Schuman et al [11] of 2.4±0.3eV, a bond dissociation energy of 1.81eV and the electron incident energy of 0.2eV, with a resultant value of 0.61eV. The kinetic energy measurement from the velocity slice map imaging recorded data (Figure 6), shows a high rise slope between 0.2eV and 1eV, with the two maximum points in similar amplitude at 0.2eV and 0.85eV.

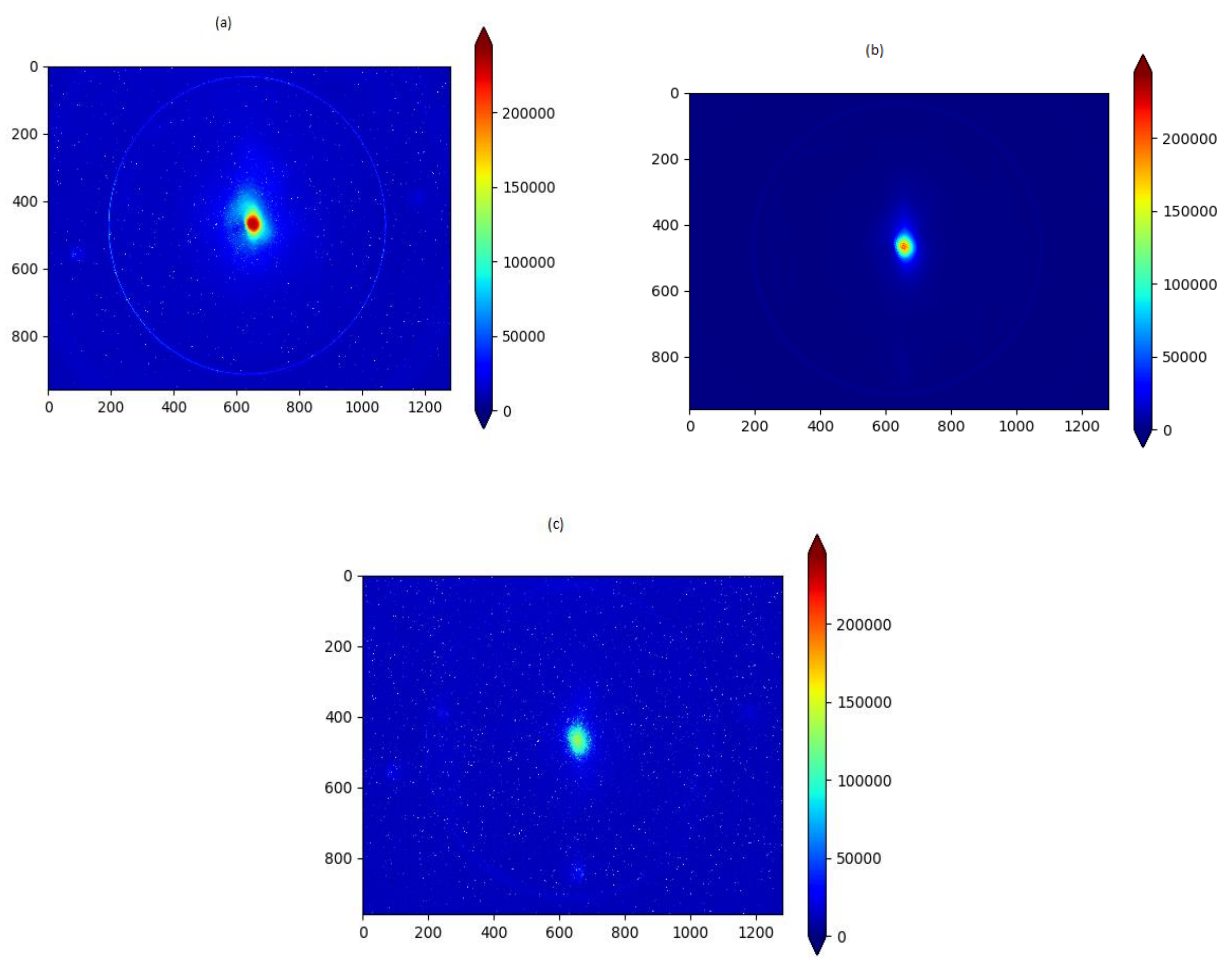

Figure 4. $Fe(CO)_5$ anions imaged by VMI CCD camera: $Fe(CO)_4^-$ (a), $Fe(CO)^-$ (b), $Fe^-$(c)

| Ions | Shuman et al AE | Shuman et al BDE | Engelking and Lineberger AE | Engelking and Lineberger Fe-C BDE | Chen et al EA |
|---|---|---|---|---|---|
| $Fe(CO)_4^-$ | 2.4 +/-0.3 | 1.81 | 1.26+/-0.02 | 1.0+/-0.3 | n/a |
| Fe(CO)$_3^-$ | 1.915+/-0.085 | 1.84 | 1.22+/-0.02 | 1.0+/-0.3 | n/a |
| Fe(CO)$_2^-$ | 1.22+/-0.02 | 1.55 | 1.8+/-0.2 | 1.4+/-0.3 | n/a |
| Fe(CO)$^-$ | 1.157+/-0.005 | 1.46 | 2.4+/-0.3 | 0.2+/-0.4 | n/a |
| Fe$^-$ | 0.151+/-0.003 | n/a | n/a | n/a | 0.153 |

Table 1. Fe(CO)$_5$ electron affinity (EA) and bond dissociation energy (BDE)

Fe(CO)$_3^-$ has a higher electron affinity in the work of Shuman et al [11], with a value of 1.915±0.085eV compared to 1.22±0.02eV reported by Engelking and Lineberger [19], with the maximum kinetic energy presenting high discrepancy between the two cited works of ~0.8eV. The fragmentation channel presents a peak at 1.3eV. Engelking and Lindeberger [19] present the appearance of a so-called "hot band" at the transition from Fe(CO)$_3^-$ to Fe(CO)$_2^-$ in the dissociation process, though the fragmentation energy of 2eV does not agree well with the VMI data reported here or the quadrupole mass spectra reported by Lacko et al [21] and cluster data from Lengyel et al [22]. The Fe$^-$ ion is presented in a separate work by Chen et al [20], who used a VMI method to determine the electron affinity by means of bombarding with Ar$^+$ atoms. The calculated kinetic energy release maximum for the Fe$^-$ ion is 9.998±0.003eV.

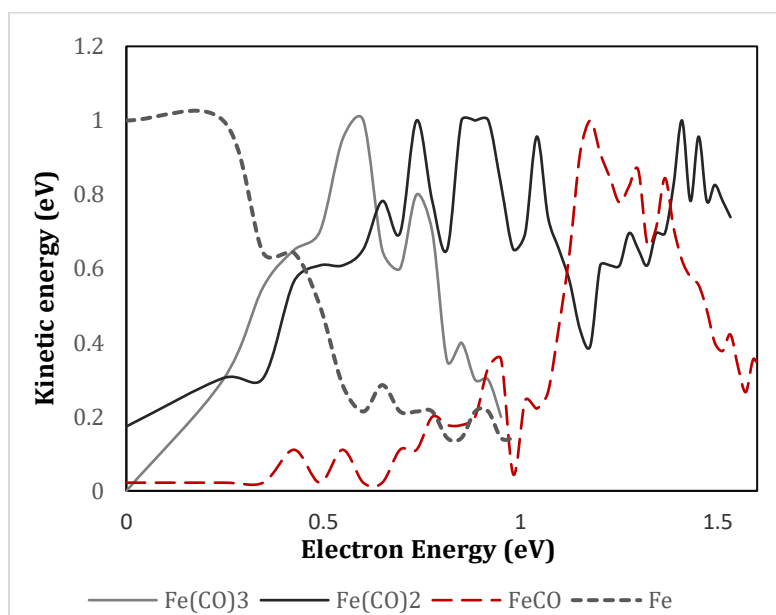

Figure 5: The angular distribution data of the negatively charged fragments, recorded with our VMI, as well as our kinetic energies show no triplet state dissociation, with the appearance of only one ring at the investigated channels.

The FeCO$^-$ is found at 8.5eV. There is a ~1eV discrepancy between the work of Shuman et al [11], with an electron affinity value of 1.157±0.005eV, and Engelking and Lineberger [19] with a value of 2.4±0.3eV, giving a kinetic energy of 8.197±0.005eV in the first case and 9.44±0.3eV, in the second case.

Engelking and Lineberger [19] present the $FeCO_2^-$ as having lower electron affinity compared to $FeCO^-$, but showing similar features. The VMI data is in good agreement with the kinetic energy calculations with values ranging between 3.87±0.02eV and 4.27±0.02eV, with electron affinities of 1.8±0.02eV and 1.22±0.02eV, Shuman et al [11].

Higher electron affinity is observed in the transition from Fe to FeCO, assigned to the bond strength with the CO, though the work of Engelking and Lineberger [19] explains this by a stabilization of d orbitals of π symmetry in the interaction with the CO π* orbitals and a destabilization of d orbitals with v symmetry by the v orbitals of CO.

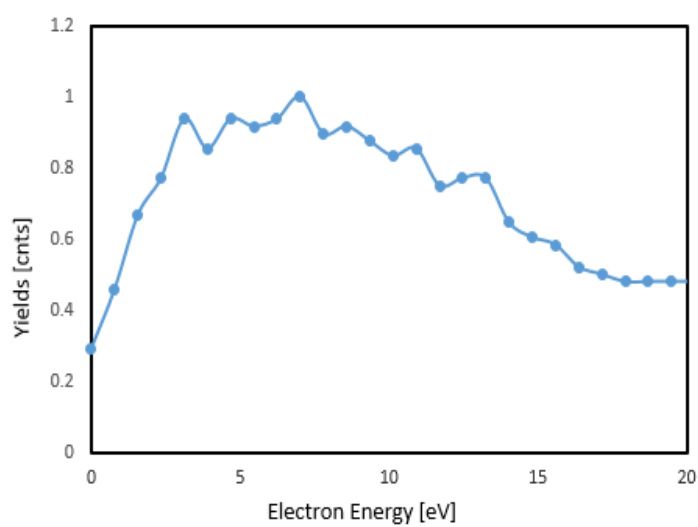

Figure 6. Kinetic energies of negative ion $Fe(CO)_4^-$

Angular distributions of the anions were derived from the VMI images but the signal to noise make these unreliable. The $Fe(CO)_4^-$ presents high symmetry, with lower amplitude around 85deg. The same behavior can be seen for $Fe(CO)_3^-$ and $Fe(CO)_4^-$ fragment  The higher fragmentation channels of $FeCO^-$ and $Fe(CO)_2^-$ present lower amplitudes up to 85deg then suggest backward cross sections. The $Fe^-$ is rather symmetric, compared to the rest of the fragments.

| Attributed channel | Peak position, eV | Peak height, ion counts | Peak width, eV |
|---|---|---|---|
| $Fe(CO)_5 + e \rightarrow Fe(CO)_5^- \rightarrow Fe(CO)_4^- + CO$ | 0.2 | 1738 | 1.49 |
| $Fe(CO)_5 + e \rightarrow Fe(CO)_5^- \rightarrow Fe(CO)_3^-$ +2CO | 1.3 | 212; 32 | 0.31;1.7;1.79;2.5 |
| $Fe(CO)_5 + e \rightarrow Fe(CO)_5^- \rightarrow Fe(CO)_2^-$ +3CO | 4.2 | 35 | 3.18; 2.9 |
| Fe(CO)$_5$ + e → Fe(CO)$^-_5$ → Fe(CO)$^-$ + 4CO | 8.5 | 1 | 5.35 |
| Fe(CO)$_5$ + e → Fe(CO)$^-_5$ → Fe$^-$ + 5CO | 9 | 8 | 4.4 |

Table 2. Experimental peak parameters and attributed channels

## 3. Simulations

### 3. 1 Numerical Setup

The VMI experimental results presented in Section 2 are complemented with scattering calculations carried out with Quantemol-N software which makes use of the UKRmol code suite [16], [23], [24]. UKRmol computes scattering cross-sections based on the R-matrix theory. In the R-matrix theory, the time-independent Schrödinger equation is solved for the system molecule + incoming electron. The incoming electron is treated as indistinguishable from the target electrons inside of a sphere which is big enough to include the molecule and its orbitals. Outside of the sphere, the scattered electron is treated separately and the solution is energy-dependent.

The true symmetry point group of Fe(CO)$_5$ is D$_{3h}$. The input symmetry point group allowed by Quantemol-N In the calculation is C2v, therefore all the orbitals and excited states found in the calculation are irreducible representations of this group, namely, [A1, B1, B2, A2]. We use the cc-pVDZ basis set for both to find the ground and excited states of the molecular target and for the scattering calculations, and the configuration Interaction quantum chemistry model is employed. The ground state energy is -1824.94 Hartree, and the configuration is 23(a1)$^2$, 11(b1)$^2$, 11(b2)$^2$, 3(a2)$^2$ which is consistent with the result obtained by Daniel et al [25]. For comparison, the NIST database gives ground state energy in the range -1825.77 to -1830.75 depending on the basis set. The highest occupied molecular orbital has b1 symmetry and energy -7.59 eV.

Scattering calculations are performed with the close coupling model, the R-matrix sphere radius is 13 Bohr. The convergence of the model was tested by increasing the complete active space (CAS) and running the calculations with different basis sets. The frozen orbitals are 22(a1)$^2$, 10(b1)$^2$, 11(b2)$^2$, 2(a2)$^2$ and the complete active space used in the calculation is 25(a1)$^2$, 11(b1)$^2$, 12(b2)$^2$, 4(a2)$^2$. Therefore there are 3 open orbitals and 4 virtual orbitals included in the calculation.

Note that the R-matrix method is less accurate at the energy above the ionization threshold. However, the calculated ionization energy is 7.59 eV which is in good agreement with Norwood et al [26].

### 3.1 Elastic electron and electron-impact excitation cross-sections

The calculation shows that the ground state is $^1$A$_1$ in the C2v point group or $^1$A$_1$' in the D3h point group. Figure 7 shows the elastic scattering cross-section for the electron energy in the range 0-10 eV.

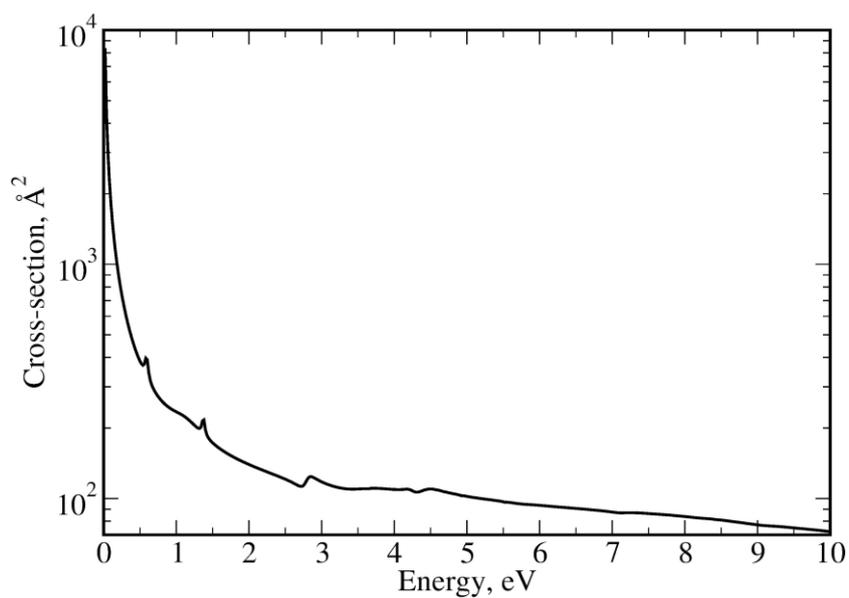

Figure 7. Electron elastic scattering cross-section for Fe(CO)$_5$

The electronic excitation of the $^1A_1$ state is characterized by the following processes:

$$Fe(CO)_5(^1A_1) + e \rightarrow Fe(CO)_5(X) + e$$

where X stands for an excited state. Energies, symmetry, and multiplicity of the first 8 excited states are given in Table 3.

| Excited state | Excitation energy, eV |
|---|---|
| $^3A2$ | 2.58 |
| $^1A2$ | 2.77 |
| $^3B2$ | 3.13 |
| $^3B1$ | 3.27 |
| $^1B2$ | 3.40 |
| $^1B1$ | 3.67 |
| $^3B1$ | 4.35 |
| $^1B1$ | 4.51 |

Table 3. Multiplicity and symmetry of the excited states in C2v and corresponding excitation energies

Figure 8 shows excitation cross-sections from the ground state to the excited states listed in Table 3. For the metal carbonyl only three allowed transitions can be seen: $^1A_1' \rightarrow {}^{1,3}E'$, $^1A_1' \rightarrow {}^{1,3}E''$ and $^1A_1' \rightarrow b^1E'$. Rubner et al [27] present the appearance of the Fe(CO)$_4$ and Fe(CO)$_3$ fragments in the frequency range of $^{1,3}E'$ and the appearance of Fe(CO)$_4$, Fe(CO)$_3$ and Fe(CO)$_2$ as a b$^1$E'. The work of Rubner et al [27] used three separate models comparing the excitation bands for each of the states: CASSCF, MR-CCI, and ACPF. For our work, only d $\rightarrow \pi_{CO}^*$ transitions states were taken into account as the only states involved in the metal-CO ligand dissociation.

The excitation cross-sections presented from our calculations are in good agreement with the data from Rubner et al [27]. It can be seen that the $^1A_1'$ excitation singlet falls in the same energy range as the dissociation excitations ~4eV for Fe(CO)$_5$ with the appearance of Fe(CO)$_4^-$, Fe(CO)$_3^-$ and Fe(CO)$_2^-$.

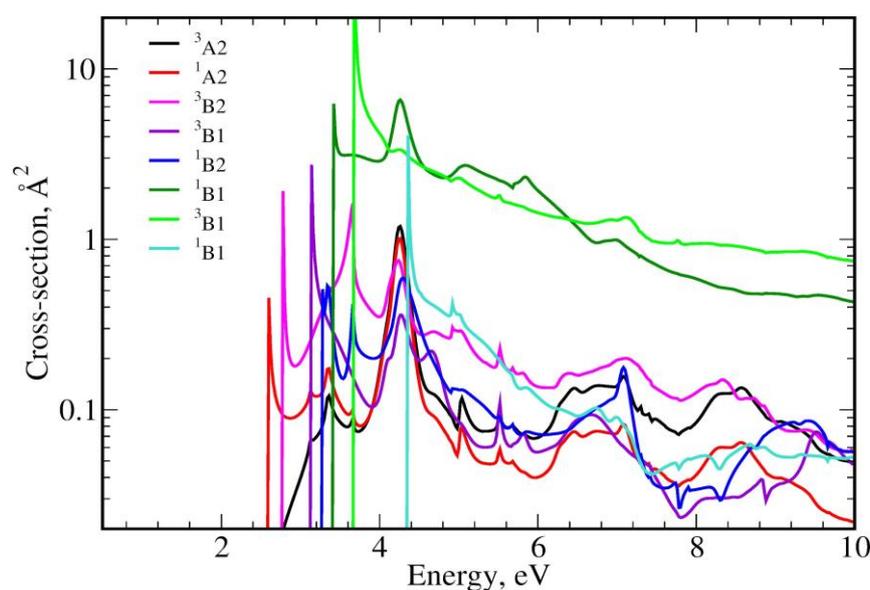

Figure 8. Electronic excitation of Fe(CO)$_5$ by electron impact from the ground state. The curves are labeled correspondingly to the excited state multiplicity and symmetry

**3.2 Dissociative electron attachment cross-sections**

Dissociative electron attachment is assumed to go via the intermediate resonance Fe(CO)$_5^-$. We have separated the energy range to a few sub-ranges of interest corresponding to different fragmentation channels. For each fragmentation channel, a specific set of parameters has to be used. Table 4 outlines the parameters used in the calculations.

In order to associate each channel with a vibrational frequency, the exact character of the modes has to be taken into account. The vibrational modes can be visualized on ChemTube3D [7]. We make an assumption that bending modes do not participate in the dissociation; if certain modes lead to simultaneous stretching of Fe - CO bonds, then these bonds will break in the DEA process.

In the dissociation process, the Fe(CO)5 molecule undergoes fast ionization from a v state to π* state, Fe(CO)$^-$

$_5$, followed by a further detachment of an electron, characterized by a thermal process Allan et al [17], or loss of one CO ligand. Allan et al [17] reports the dissociation of the Fe(CO)$_5^-$ into Fe(CO)$_4^-$ + CO as plausible at two peak energies, 0.08eV and 0.25eV electron energy. From the mass spectrum, the peaks corresponding to this fragmentation channel are at 0.08eV and 0.26eV. From our VMI experiments, we found the fragmentation channel leading to the formation of Fe(CO)$_4^-$ at 0.2eV incident energy, close to the data reported by them with a bond dissociation energy of 1.81eV.

| DEA fragmentation channel | Energy range, eV | Vibrational frequency, cm$^{-1}$, mode | Electron affinity, eV | Bond dissociation energy, eV |
|---|---|---|---|---|
| Fe(CO)$_5$ + e- → Fe(CO)$_4^-$ + CO | 0 - 1 | 2093.4 (E') | 2.4±0.3 | 1.81 |
| Fe(CO)$_5$ + e- → Fe(CO)$_3^-$ + 2(CO) | 1 - 2.5 | 2118.85 (A1') | 1.915±0.085 | 1.84 |
| Fe(CO)$_5$ + e- → Fe(CO)$_2^-$ + 3(CO) | 2.5-6 | 2093.4 (E') | 1.22±0.02 | 1.55 |
| Fe(CO)$_5$ + e- → Fe(CO)$^-$ + 4(CO) | 4-10 | 2120.24 (A2'') | 1.157±0.005 | 1.46 |
| Fe(CO)$_5$ + e- → Fe$^-$ + 4(CO) | 6-10 | 2192.65 (A1') | 0.151±0.003 | |

Table 4. Dissociative electron attachment parameters for each fragmentation channel. Electron affinity and bond breaking energies are from [11]

The CO-stretch excitation cross-section was found at 0.66eV that might correspond to the 0.2eV Fe(CO)$_4^-$ band or to the 1eV width wide Fe(CO)$_4^-$ shoulder starting at 0.8eV. The CO-stretch for 0.7eV band has been reported in the literature to be due to the same π* resonance as the 1.3eV band, as Feshbach resonances are usually 0 - 0.4eV lower than the parent electronically excited state. Our calculations took into account an electron affinity of 1.2 - 2.4eV and a vibrational frequency in the range 1000 – 2100cm$^{-1}$ with a user-defined basis set based on cc-pVDZ with C2υ geometry.

The resonances were found with the R-matrix routine RESON [29]. RESON returns all the resonances found in the process for each energy range. We calculate DEA cross-sections using specific dissociation energy, electron affinity and the effective mass corresponding to the fragmentation channel.

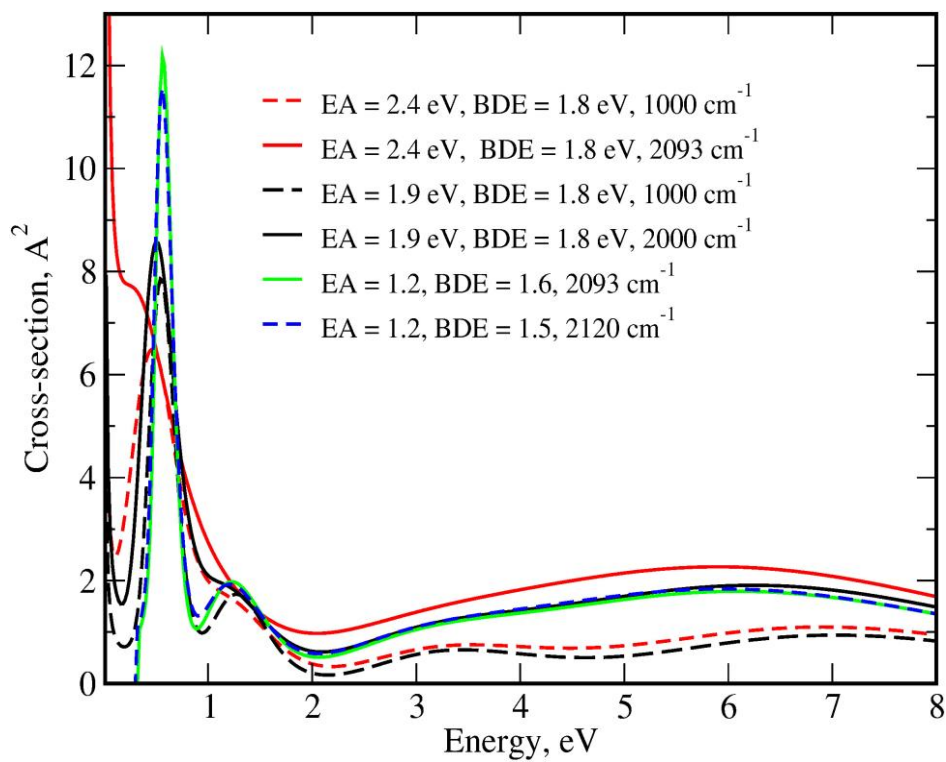

Figure 9. Dissociative electron attachment cross-section as a function of electron affinity (EA), bond dissociation energy (BDE), and vibrational frequency

The total cross sections calculated from the DEA process are presented in Figure 9. While the positions of the peaks are generally determined by the positions of the resonances and not by the parameters of the DEA model, they may slightly shift or get wider, and the relative heights also change as a function of the electron affinity, bond dissociation energy, and the vibrational frequency (to a lesser extent) which are, in turn, functions of the fragmentation channel. The main changes can be observed at low energies. Red curves in Figure 9 correspond to the electron affinity and the dissociation energy leading to the formation of $Fe(CO)_4^-$. For comparison, we show the cross-section for 2 vibrational frequencies: 1000 $cm^{-1}$ and 2000 $cm^{-1}$. One can see that the peak at 0 becomes more prominent if the vibrational frequency is lower. These curves describe DEA below 1 eV.

The black curves correspond to the second fragmentation channel and the formation of $Fe(CO)_3^-$. The relevant energy region is 1-2.5eV and for this level, the lower vibrational frequencies lead to the prevalence of the peaks at lower energies. Comparing the red and the black sets of curves, one can see lower electron affinity leads to a more pronounced peak (or better-separated peaks).

At higher energies and for the parameters corresponding to $Fe(CO)_2^-$ and $Fe(CO)^-$, the specific peak structure cannot be resolved, although one can see the signatures of the resonances between 3 and 4 eV and 6 and 7 eV.

A peak at 0.6eV with the highest intensity is observed in the range of 12.5 x $10^{-20}$ $m^2$ corresponding to $Fe(CO)_4^-$, followed by a second peak corresponding to $Fe(CO)_3^-$ around 1.4eV with cross-sections in the range of 5.5 x $10^{-20}$ $m^2$ and a $Fe(CO)_2^-$ peak between 4eV and 5ev, the $Fe^-$ and $FeCO^-$ peaks are overlapping and further partial calculations are run separately for each of them. The estimation method for partial cross-sections in the DEA process is not a replacement for experimental data and calculations, but a cross-check method.

**CONCLUSIONS**

In this paper we report new results on DEA studies of the Fe(CO)$_5$ molecule. The complete fragmentation stripping off all carbonyl groups makes it a great candidate for FEBID deposition. We have also presented theoretical calculations of the elastic, excitation and DEA cross sections using the Quantemol-N program. We hope this data may be used in further simulations of Fe(CO)$_5$ within a FEBID processing system.

**Acknowledgements.**
MP recognizes receipt of a Marie Curie Early Career Award from the ELENA ITN which received funding from the European Union's Horizon 2020 research and innovation program under the Marie Skłodowska-Curie grant agreement No 722149. The authors wish to acknowledge the collection of DEA ion spectra by P Papp and M Danko at Comenius University. Slovakia (shown in figure 3).

**References**

[1] De Teresa J. M., Fernandez - Pacheco A. (2014), Present and future applications of magnetic nanostructures grown by FEBID, Appl Physics A (2014) 117: 1645-1658

[2] van Dorp W. F., Hagen C. W. (2008), A critical literature review of focused electron beam induced deposition, J Appl Phys 104, 081301

[3] Gavanin M., Wanzenboeck H.D., Belic D., Shawrav M.M., Persson A., Gunnarsson K., Svedlindh P., Bertagnolli E. (2014) Magnetic force microscopy study of shape engineered FEBID iron nanostructures, Phys. Status Solidi A 211, No. 2, 368-374

[4] Huth M., Porrati F., Dobrovolskiy O.V. (2018) Focused electron beam induced deposition meets materials science, Microelectronic Engineering 185-186, 2018, 9–28

[5] Lukasczyk T, Schirmer M, Steinrück HP, Marbach H., (2008) Electron-beam-induced deposition in ultrahigh vacuum: lithographic fabrication of clean iron nanostructures. Small. 4(6):841- 846

[6] Thorman R.M., Kumar R.T.P., Fairbrother D.H., Ingólfsson, (2015) The role of low-energy electrons in focused electron beam induced deposition: four case studies of representative precursors, Beilstein J. Nanotechnol., 6, 1904–1926

[7] Kumar R., Unlu I., Barth S., Ingólfsson O., Fairbrother D.H. (2018) Electron Induced Surface Reactions of HFeCo$_3$(CO)$_{12}$, a Bimetallic Precursor for Focused Electron Beam Induced Deposition (FEBID), J. Phys. Chem. C, 122 2648–2660

[8] Unlu I., Spencer J., Johnson K.R., Thorman R., Ingólfsson O., McElwee-White L., Fairbrother D.H. (2018) Electron induced surface reactions of ($\eta^5$-$C_5H_5$)Fe(CO)$_2$Mn(CO)$_5$, a potential heterobimetallic precursor for focused electron beam induced deposition (FEBID), Phys. Chem. Chem. Phys. 20 7862-7874

[9] Sha L., Porcel E., Remita H., Marco S., Réfrégiers M., Dutertre M., Confalonieri F., Lancombe S. (2017) Platinum nanoparticles: an exquisite tool to overcome radioresistance, Cancer Nanotechnology 8.1 : 4

[10] Verkhovtsev A., Traore A., Muñoz A., Blanco F., García G. (2017) Modeling secondary particle tracks generated by intermediate- and low-energy protons in water with the Low-Energy Particle Track Simulation code, Radiation Physics and Chemistry 130 371-378

[11] Shuman N. S., Miller T. M., Friedman J. F., Viggiano A. A. (2013) Electron Attachment to Fe(CO)$_n$ ($n$ = 0–5), J Phys Chem A2013, 117, 1102-1109

[12] Lengyel J., Fedor J., Farnik M. (2016) Ligand Stabilization and Charge Transfer in Dissociative Ionization of Fe(CO)$_5$ Aggregates, J Phys Chem C 2016, 120, 17810-17816

[13] Kunnus K., Josefsson I., Rajkovic I., Schreck S., Quevedo W., Beye M., Weniger C., Grübel S., Scholz M., Nordlund D., Zhang W., Hartsock R. W., Gaffney K. J., Schlotter W. F., Turner J. J., Kennedy B., Hennies F., de Groot F. M. F., Techert S., Odelius M., Wernet Ph, Föhlisch A. (2016) Identification of the dominant photochemical pathways and mechanistic insights to the ultrafast ligand exchange of Fe(CO)5 to Fe(CO)4EtOH, Struct Dyn 3, 043204

[14] Aiswaryalakshmi P., Mani D., Arunan E. (2013) Fe as Hydrogen/Halogen Bond Acceptor in Square Pyramidal Fe(CO)$_5$, Inorg Chem, 52, 9153-9161

[15] Wernet Ph., Kunnus K., Josefsson I., RRajkovic I., Quevado W., Beye M., Schreck S., Grübel S., Scholz M., Nordlund D., Zhang W., Hartsock R. W., Schlotter W. F., Turner J. J., Kennedy B., Hennies F., de Groot F. M. F., Gaffney K. J., Techert S., Odelius M., Föhlisch A. (2015) Orbital-specific mapping of the ligand exchange dynamics of Fe(CO)$_5$ in solution, Nature, Vol 520, 78-81

[16] Munro J., Harrison S., Fujimoto M., Tennyson J., (2012) A dissociative electron attachment cross-section estimator, J Phys: Conf Ser 388 012013

[17] Allan M., Lacko M., Papp P., Matejčík Š., Zlatar M., Fabrikant I. I., Kočišek J., Fedor J. (2018), Dissociative electron attachment and electronic excitation in Fe(CO)$_5$, Phys Chem Chem Phys, 20, 11692

[18] Lengyel J., Papp P., Matejčík Š., Kočišek J., Fárník M., Fedor J. (2017), Suppression of low-energy dissociative electron attachment in Fe(CO)$_5$ upon clustering, Beilstein J. Nanotechnology, 2017, 8, 2200-2207

[19] Engelking P. C., Lineberger W. C. (1979) Laser Photoelectron Spectrometry of the Negative Ions of Iron and Iron Carbonyls. Electron Affinity Determination for the Series Fe(CO)$_n$ = 0, 1, 2, 3, 4, Journal of the American Chemical Society, 101:19

[20] Chen X., Luo Z., Li J., Ning C. (2016) Accurate Electron Affinity of Iron and Fine Structures of Negative Iron ions, Scientific Reports 6, 24996

[21] Lacko M., Papp P., Wnorowski K., Matejčík Š. (2015) Electron-induced ionization and dissociative ionization of iron pentacarbonyl molecules, Eur Phys JD, 69:84

[22] Lengyel J., Kočišek J., Fedor J. (2016) Self-Scavenging of Electrons in Fe(CO)$_5$ Aggregates Deposited on Argon Nanoparticles, J Phys Chem C, 120, 7397-7402

[23] Hamilton J. R., Tennyson J., Huang S., Kushner M. J., (2017) Calculated cross sections for electron collisions with NF3 , NF2 and NF with applications to remote plasma sources, Plasma Sources Sci Technol 26, 065010

[24] Song M.–Y., Yoon J.-S., Cho H., Karwasz G. P., Kokoouline V., Nakamura Y., Tennyson J., (2015) Cross Sections for Electron Collisions with Methane, J. Phys. Chem. Ref. Data, Vol. 44, No. 2

[25] Norwood K., Ali A., Flesh G.D., Ng C.Y (1990) A Photoelectron-Photoion Coincidence Study of Fe(CO)$_5$, J. Am. Chem. Soc., 1990, 112, 7502

[26] Clayden J., Greeves N., Warren S. (2008) ChemTube3D - Interactive 3D Organic Reaction Mechanisms, http://www.chemtube3d.com/ . Accessed November 2019

[27] Rubner O., Engel V., Hachey M. R., Daniel C. (1999) A CASSCF/MR-CCI study of the excited states of Fe(CO)$_5$, Chemical Physics Letters 302 (1999), 481-494

[28] Tennyson J., Noble C. J. (1984) RESON - A PROGRAM FOR THE DETECTION AND FITHNG OF BREIT-WIGNER RESONANCES, Comput. Phys. Commun. 33, 421–424

[29] Szymańska E., Prabhudesai V. S., Mason N. J., Krishnakumar E., (2013) Dissociative electron attachment to acetaldehyde, $CH_3CHO$. A laboratory study using the velocity map imaging technique, Phys Chem Chem Phys, 2013, 15, 998-1005

[30] Lengyel J., Fedor J., Farnik M. (2016) Ligand Stabilization and Charge Transfer in Dissociative Ionization of Fe(CO)$_5$ Aggregates, J Phys Chem C 2016, 120, 17810-17816

[31] Trushin S. A., Fuss W., Kompa K. L., Schmid W. E. (2000) Femtosecond Dynamics of Fe(CO)$_5$ Photodissociation at 267 nm Studied by Transient Ionization, J Phys Chem A 2000, 104, 1997-2006

[32] Massey S., albas A. D., Sanche L. (2015) Role of Low-Energy Electrons (<35 eV) in the Degradation of Fe(CO)$_5$ for Focused Electron Beam Induced Deposition Applications: Study by Electron Stimulated Desorption of Negative and Positive Ions, J Phys Chem C, 119, 12708-12719

[33] Shuman N. S., Miller T. M., Friedman J. F., Viggiano A. A. (2013) Electron Attachment to Fe(CO)$_n$ ($n$ = 0–5), J Phys Chem A, 117, 1102-1109

[34] Willock D. (2009) Molecular Symmetry, book, Wiley, pp127

[35] Sathyanarayana N. (2004) Vibrational Spectroscopy: Theory and Applications, book, New Age International, pp 368

[36] Squires R. R., Wang D., Sunderlin L. S. (1992) Metal Carbonyl Bond Strengths in Fe(CO)$_n^-$ and Ni(CO)$_n^-$, J. Am. Chem. Soc., 114, 2788-2796

[37] Snee P. T., Payne C. K., Mebane S. D., Kotz K. T., Harris C. B. (2001) Dynamics of Photosubstitution Reactions of Fe(CO)$_5$: An Ultrafast Infrared Study of High Spin Reactivity, J. Am. Chem. Soc., Vol. 123, No. 28

[38] Besora M., Carreón-Macedo J.-L., Cowan A. J., George M. W., Harvey J. N., Portius P., Ronayne K. L., Sun X. – Z., Towrie M. (2009) A Combined Theoretical and Experimental Study on the Role of Spin States in the Chemistry of Fe(CO)$_5$ Photoproducts, J. Am. Chem. Soc., 131, 3583-3592

[39] Daniel C., Benard M., Dedieu A., Wiest R., Veillard A. (1984) Theoretical Aspects of the Photochemistry of Organometallics. 3. Potential Energy Curves for the Photodissociation of Fe(CO)$_5$, The Journal of Physical Chemistry, 88(21), 4805–4811

[40] Nandi D., Prabhudesai V. S., Krishnakumar E., (2005) Velocity slice imaging for dissociative electron attachment, Review of Scientific Instruments 76, 053107

[41] Brigg W. J., Tennyson J., Plummer M., (2014) R-matrix calculations of low-energy electron collisions with methane, J. Phys. B: At. Mol. Opt. Phys. 47 185203

[42] Goswami B., Naghma R., Antony B. (2014) Electron scattering from germanium tetrafluoride, RSC Advances 4 63817

[43] Sieglaff D. R., Rejoub R., Lindsay B. G., Stebbings R. F. (2001) Absolute partial cross sections for electron-impact ionization of CF4 from threshold to 1000eV, J Phys B: At Mol Opt Phys 34, 1289-1297

[44] Bhargava Ram N., Prabhudesai V. S., Krishnakumar E., (2012) Dynamics of the dissociative electron attachment in $H_2O$ and $D_2O$: The A1 resonance and axial recoil approximation, J Chem Sci Vol 124, No 1, pp 271-279

[45] Bhargava Ram N., Krishnakumar E. (2012) Dissociative electron attachment resonances in ammonia: a velocity slice imaging based study, J Chem Phys 136, 164308

[46] Bhargava Ram, Krishnakumar E. (2011) Dissociative electron attachment to methane probed using velocity slice imaging, Chemical Physics Letters, 511 (2011), 22-27

[47] Gope K., Tadsare V., Prabhudesai V. S., Mason N. J., Krishnakumar E., (2016) Negative ion resonances in carbon monoxide, Eur Phys JD (2016) 70:134

[48] Bjarnason E., Ómarsson B., Engmann S., Ómarsson F. H., Ingólfsson (2014) Dissociative electron attachment to titanium tetrachloride and titanium tetraisopropoxide, Eur Phys JD 68:101

**Appendix I. The XYZ atomic coordinates used in the calculation**

| Atom | X, Å | Y, Å | Z, Å |
|---|---|---|---|
| Fe | 0 | 0 | 0 |
| C | 0 | -2.0402 | 0 |
| C | 0 | 0 | 1.8462 |
| C | 1.5989 | 0 | -0.9231 |
| C | -1.5989 | 0 | -0.9231 |
| C | 0 | 2.0402 | 0 |
| O | 0 | -3.1438 | 0 |
| O | 0 | 0 | 2.9619 |
| **O** | 2.5651 | 0 | -1.4810 |
| **O** | -2.5651 | 0 | -1.4810 |
| **O** | 0 | 3.1438 | 0 |

Table 1. Atomic coordinates of Fe(CO)$_5$